\documentclass[aps,prb,twocolumn,showpacs]{revtex4}
\usepackage{graphicx}

\begin{document}

\title{Superconductivity at 15.6 K in Calcium-doped Tb$_{1-x}$Ca$_x$FeAsO: the structure requirement for achieving superconductivity in the hole-doped 1111 phase}

\author{Gang Mu, Bin Zeng, Peng Cheng, Xiyu Zhu, Fei Han, Bing Shen, and Hai-Hu Wen}\email{hhwen@aphy.iphy.ac.cn}

\affiliation{National Laboratory for Superconductivity, Institute of
Physics and Beijing National Laboratory for Condensed Matter
Physics, Chinese Academy of Sciences, P. O. Box 603, Beijing 100190,
People's Republic of China}

\begin{abstract}
Superconductivity at about 15.6 K was achieved in
Tb$_{1-x}$Ca$_x$FeAsO by partially substituting Tb$^{3+}$ with
Ca$^{2+}$ in the nominal doping region $x = 0.40 \sim 0.50$. A
detailed investigation was carried out in a typical sample with
doping level of $x$ = 0.44. The upper critical field of this sample
was estimated to be 77 Tesla from the magnetic field dependent
resistivity data. The domination of hole-like charge carriers in the
low-temperature region was confirmed by Hall effect measurements.
The comparison between the calcium-doped sample
Pr$_{1-x}$Ca$_x$FeAsO (non-superconductive) and the Strontium-doped
sample Pr$_{1-x}$Sr$_x$FeAsO (superconductive) suggests that a lager
ion radius of the doped alkaline-earth element compared with that of
the rare-earth element may be a necessary requirement for achieving
superconductivity in the hole-doped 1111 phase.
\end{abstract}

 \maketitle

\section{Introduction}

The discovery of superconductivity in iron pnictides have generated
enormous interests in the community of condensed matter
physics.\cite{Kamihara2008} Up to date, the iron pnictide
superconductors have developed into several families with different
structures, which were abbreviated as the 1111 phase (including the
oxy-arsenide\cite{Kamihara2008} and fluorine-arsenide\cite{SrF}),
122 phase,\cite{Rotter,CWCh} 111
phase,\cite{LiFeAs,LiFeAsChu,LiFeAsUK} 11 phase,\cite{FeSe} 42622
phase,\cite{42622} and so on. It seems that each phase with
different structure has a unique superconducting transition
temperature $T_c$. As for the 1111 phase, most of the discovered
superconductors are characterized as electron-doped ones
\cite{Pr52K,CP,WangC,Mandrus,CaoGH}, while the hole-doped
superconductors were only reported in the strontium-doped
Ln$_{1-x}$Sr$_{x}$FeAsO (Ln = La, Pr, Nd).\cite{WenEPL,LaSr2,PrSr,
NdSr} The hole-doped superconductivity in 1111 phase by substituting
other ion-dopants with valence of "+2", such as barium or calcium,
seems quite difficult to be achieved, at least in many of the
rare-earth based systems. Obviously, it is important to carry out
more explorations in this direction in order to extend the family of
the hole-doped superconductors in 1111 phase. And it is also
significant to investigate the factors which govern the electronic
properties (superconducting or non-superconducting) in the
hole-doped side based on the 1111 phase.

In this paper we report a new hole-doped superconductor in the 1111
phase, calcium-doped Tb$_{1-x}$Ca$_x$FeAsO, with the maximum
superconducting transition temperature of 15.6 K (95\% $\rho_n$). It
is found that superconductivity appears in the nominal doping region
$x = 0.40 \sim 0.50$. The physical properties of a selected sample
with $x$ = 0.44 were investigated in depth. We estimated the upper
critical field of this sample to be 77 Tesla based on the
Werthamer-Helfand-Hohenberg (WHH) formula.\cite{WHH}  The conducting
charge carriers in this sample were characterized to be hole type in
a wide low-temperature region by the Hall effect measurements.
Meanwhile, we have also successfully synthesized calcium-doped
Pr$_{1-x}$Ca$_x$FeAsO, which also displays hole-type charge carriers
in low-temperature region but doesn't superconduct at all. We
attribute this different behavior to the sensitive electronic
response to the relative radii of the doped ions compared with that
of the rare-earth ions.

\section{Experimental Details}

The Tb$_{1-x}$Ca$_x$FeAsO samples were prepared using a two-step
solid state reaction method. In the first step, TbAs and CaAs were
prepared by reacting Tb flakes (purity 99.99\%), Ca flakes (purity
99.9\%) and As grains (purity 99.99\%) at 500 $^o$C for 10 hours and
then 700 $^o$C for 16 hours. They were sealed in an evacuated quartz
tube when reacting. Then the resultant precursors were thoroughly
grounded together with Fe powder (purity 99.95\%) and Fe$_2$O$_3$
powder (purity 99.5\%) in stoichiometry as given by the formula
Tb$_{1-x}$Ca$_x$FeAsO. All the weighing and mixing procedures were
performed in a glove box with a protective argon atmosphere. Then
the mixtures were pressed into pellets and sealed in an evacuated
quartz tube. The materials were heated up to 1150-1170 $^o$C with a
rate of 120 $^o$C/hr and maintained for 40 hours. Then a cooling
procedure was followed. After that, we can get the superconducting
polycrystalline samples. The process of preparing
Pr$_{1-x}$Ca$_x$FeAsO samples is quite similar to that of
Tb$_{1-x}$Ca$_x$FeAsO.

The x-ray diffraction (XRD) measurements of our samples were carried
out by a $Mac$-$Science$ MXP18A-HF equipment with Cu-K$_\alpha$
radiation. The ac susceptibility of the samples were measured on the
Maglab-12T (Oxford) with an ac field of 0.1 Oe and a frequency of
333 Hz. The resistance and Hall effect measurements were done using
a six-probe technique on the Quantum Design instrument physical
property measurement system (PPMS) with magnetic fields up to 9 T.
The current direction was changed for measuring each point in order
to remove the contacting thermal power. The temperature
stabilization was better than 0.1\% and the resolution of the
voltmeter was better than 10 nV.

\begin{figure}
\includegraphics[width=9cm]{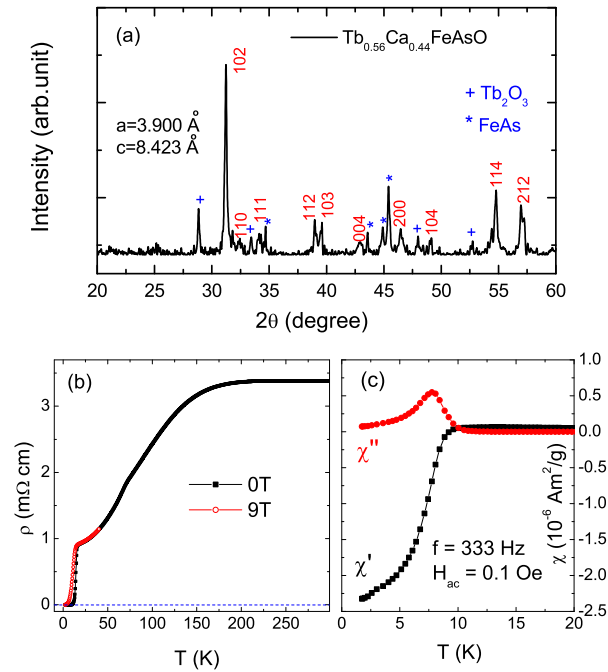}
\caption{(Color online) (a) X-ray diffraction pattern for the sample
Tb$_{0.56}$Ca$_{0.44}$FeAsO. All the main peaks can be indexed to
the tetragonal ZrCuSiAs-type structure. The peaks from the
impurities are precisely indexed to Tb$_2$O$_3$ and FeAs. (b)
Temperature dependence of resistivity for the
Tb$_{0.56}$Ca$_{0.44}$FeAsO sample under two different fields 0 T
and 9 T. The data under 0 T is shown up to 300 K. (c) The ac
susceptibility data measured with $f = 333$ Hz and $H_{ac} = 0.1$ Oe
.} \label{fig1}
\end{figure}

\section{Experimental data and discussion}

\subsection{Sample characterization for Tb$_{0.56}$Ca$_{0.44}$FeAsO}

The x-ray diffraction pattern for the sample Tb$_{1-x}$Ca$_x$FeAsO
with the nominal doping level of $x$ = 0.44 is shown in Fig. 1(a).
It is clear that all the main peaks can be indexed to the 1111 phase
with the tetragonal ZrCuSiAs-type structure.\cite{ZrCuSiAs} The main
impurity phases were identified to be Tb$_2$O$_3$ and FeAs, which
are all not superconducting in the measuring temperature. By using
the software Fullprof, we can determine the lattice constants as $a
= 3.900$ $\AA$ and $c = 8.423$ $\AA$ for this sample. By comparing
with the lattice constants of the parent phase TbFeAsO ($a = 3.898$
$\AA$, $c = 8.404$ $\AA$) reported by other group,\cite{JieY} we
find that the $a$-axis lattice constant in the present sample is
slightly larger than that of the parent phase, while the expansion
along the $c$-axis direction is more distinct. In fact, the similar
tendency has been observed in other hole-doped systems in the 1111
phase.\cite{LaSr2,PrSr} This indicates that the calcium atoms go
into the crystal lattice of the TbFeAsO system because the radius of
Ca$^{2+}$ is larger than that of Tb$^{3+}$ (see Fig. 7).

In Fig. 1(b) we present a typical set of resistive data for the same
sample Tb$_{0.56}$Ca$_{0.44}$FeAsO under 0 T and 9 T. The data under
0 T is shown up to 300 K. A clear superconducting transition can be
seen in the low temperature region. Taking a criterion of 95\%
$\rho_n$, the onset transition temperature is determined to be 15.6
K. A magnetic field of 9 T only suppresses the onset transition
temperature about 1.6 K, indicating a rather high upper critical
field in our sample. In the high temperature region, the resistivity
anomaly coming from the antiferromagnetic (AF) or structural
transition has been suppressed and a flattening feature was observed
clearly. The similar behavior has been observed in other hole-doped
1111 systems Ln$_{1-x}$Sr$_x$FeAsO (Ln = La, Pr,
Nd).\cite{WenEPL,LaSr2,PrSr,NdSr} Figure 1(c) shows the ac
susceptibility data measured with $f = 333$ Hz and $H_{ac} = 0.1$
Oe. A rough estimate from the diamagnetic signal shows that the
superconducting volume fraction of the present sample is beyond
50\%, confirming the bulk superconductivity in our samples. The
onset critical temperature by magnetic measurements is roughly
corresponding to the zero-resistance temperature.

\begin{figure}
\includegraphics[width=8.5cm]{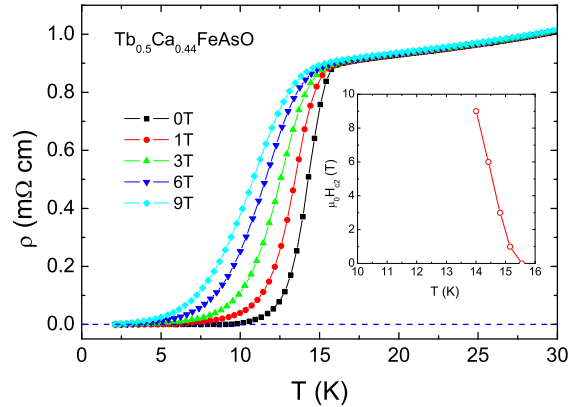}
\caption{(Color online) Temperature dependence of resistivity for
Tb$_{0.56}$Ca$_{0.44}$FeAsO near the superconducting transition
under different magnetic fields. The onset transition temperature
defined by 95\%$\rho_n$ shifts with the magnetic field slowly.
Inset: phase diagram derived from the resistive transition curves.}
\label{fig2}
\end{figure}

\begin{figure}
\includegraphics[width=9cm]{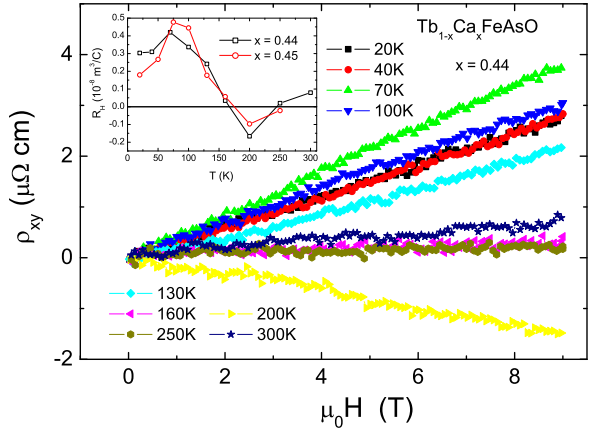}
\caption{(Color online) Hall effect measurements for the samples
Tb$_{1-x}$Ca$_{x}$FeAsO. The main frame shows the field dependence
of the Hall resistivity $\rho_{xy}$ at different temperatures for
the sample with $x$ = 0.44. Inset: Temperature dependence of the
Hall coefficient $R_H$ for two samples with $x$ = 0.44 and 0.45.}
\label{fig3}
\end{figure}

\subsection{Upper critical field for Tb$_{0.56}$Ca$_{0.44}$FeAsO}

We attempted to estimate the upper critical field of the sample
Tb$_{0.56}$Ca$_{0.44}$FeAsO from the resistivity data. Temperature
dependence of resistivity under different magnetic fields is shown
in the main frame of Fig. 2. It is found that the onset transition
point, which reflects mainly the upper critical field in the
configuration of H$\|$ab-plane, shifts more slowly than the zero
resistivity point to low temperatures under fields. The
magnetoresistance in the normal state is found to be quite small. We
take a criterion of 95\%$\rho_n$ to determine the onset transition
points under different fields, which are represented by the red open
circles in the inset of Fig. 2. From these data we can determine the
slope of $H_{c2}(T)$ near $T_c$, $dH_{c2}/dT|_{T_c} \approx -7.1$
T/K. By using the WHH formula\cite{WHH} the value of zero
temperature upper critical field $H_{c2}(0)$ can be estimated
through:
\begin{equation}
H_{c2}(0)=-0.693T_c(\frac{dH_{c2}}{dT})|_{T_c}. \label{eq:1}
\end{equation}
Taking $T_c$= 15.6 K, we get $H_{c2}(0) \approx 77$ T. Regarding the
relatively low value of $T_c$=15.6 K in the present sample, this
value of upper critical field $H_{c2}(0)$ is actually quite high.

Actually, in the strontium-doped Ln$_{1-x}$Sr$_x$FeAsO (Ln = La,
Pr), the rather high $H_{c2}(0)$ and large slope
$dH_{c2}(T)/dT|_{T_c}$ ($\sim4$ T/K) have been observed when
comparing with the F-doped LaFeAsO sample, which was attributed to
higher quasiparticle density of states (DOS) near the Fermi level in
the hole-doped samples.\cite{PrSr} Surprisingly, the slope
$dH_{c2}(T)/dT|_{T_c}$ found here is even larger than that of the
strontium-doped samples. The essential physical mechanism for this
behavior may still need more investigation in this system, including
that from the theoretical side.

\subsection{Hall effect of Tb$_{1-x}$Ca$_{x}$FeAsO}

It is known that Hall effect measurement is a useful tool to
investigate the information of charge carriers and the band
structure. For a conventional metal with Fermi liquid feature, the
Hall coefficient is almost independent of temperature. However, this
situation is changed for a multiband material\cite{HY} or a sample
with non-Fermi liquid behavior, such as the cuprate
superconductors.\cite{Ong} To examine the type of the conducting
carriers, we measured the Hall effect of the samples
Tb$_{1-x}$Ca$_{x}$FeAsO. The main frame of Fig. 3 shows the magnetic
field dependence of Hall resistivity ($\rho_{xy}$) at different
temperatures for the sample with $x$ = 0.44. In the experiment
$\rho_{xy}$ was taken as $\rho_{xy}$ = [$\rho$(+H) - $\rho$(-H)]/2
at each point to eliminate the effect of the misaligned Hall
electrodes. We can see that all curves in Fig. 3 have good linearity
versus the magnetic field. Moreover, $\rho_{xy}$ is positive at all
temperatures below 160 K giving a positive Hall coefficient $R_H =
\rho_{xy}/H$, which actually indicates that hole-type charge
carriers dominate the conduction below 160 K in the present sample.

The temperature dependence of $R_H$ for two samples with $x$ = 0.44
and 0.45 is shown in the inset of Fig. 3. One can see that the
evolution of $R_H$ with temperature are quite similar for the two
samples, indicating the reliability of the Hall data. The hump
feature in low temperature region is quite similar to that observed
in strontium-doped Ln$_{1-x}$Sr$_{x}$FeAsO (Ln = La, Pr)
samples.\cite{WenEPL,LaSr2,PrSr} However, there is still some
differences obviously. Firstly, the Hall coefficient $R_H$ changes
its sign at about 160 K which is remarkably lower than that observed
in the strontium-doped systems ($\sim250$ K). This character seems
to be quite common in the calcium-doped 1111 phase because the sign
changing of $R_H$ was also found to occur at about 160 K in
Pr$_{1-x}$Ca$_{x}$FeAsO (see Fig. 5) and Nd$_{1-x}$Ca$_{x}$FeAsO
(data not shown here). Secondly, the negative $R_H$ at about 200 K
has a rather large absolute value. This feature seems to be unique
in the calcium-doped superconducting samples, since it can't be
observed in Ln$_{1-x}$Sr$_{x}$FeAsO (Ln = La, Pr) or
Pr$_{1-x}$Ca$_{x}$FeAsO. Assuming a simple two-band scenario with
different types of carriers, we can express the Hall coefficient
$R_H$ in the low-field limit as
\begin{equation}
R_H=\frac{\sigma_1 \mu_1+\sigma_2
\mu_2}{(\sigma_1+\sigma_2)^2},\label{eq:2}
\end{equation}
where $\sigma_i$ and $\mu_i$ are the conductivity and the mobility
of the $i^{th}$ band, respectively. They are determined by the
charge-carrier density and scattering rate of each band. We
attribute the strong and complicated temperature dependence of $R_H$
in the present system to the competing effect of the scattering rate
as well as the charge-carrier density in different bands.

\begin{figure}
\includegraphics[width=9cm]{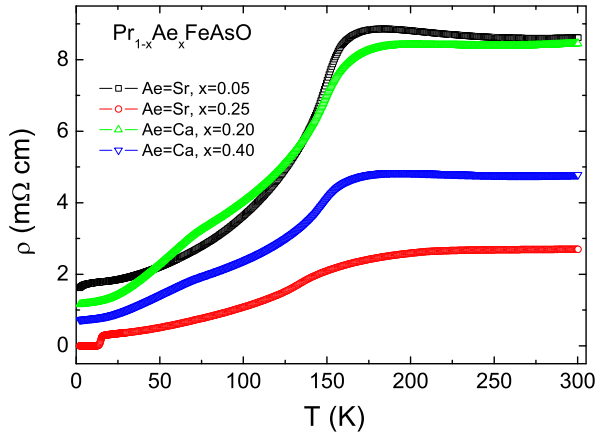}
\caption{(Color online) Temperature dependence of resistivity for
two calcium-doped samples Pr$_{1-x}$Ca$_{x}$FeAsO with $x$ = 0.20
and 0.40, along with two strontium-doped samples
Pr$_{1-x}$Sr$_{x}$FeAsO with $x$ = 0.05 and 0.25 for comparison. It
is clear that the behavior of the calcium-doped samples is between
that of the two strontium-doped samples in high temperature region.}
\label{fig4}
\end{figure}

\begin{figure}
\includegraphics[width=9cm]{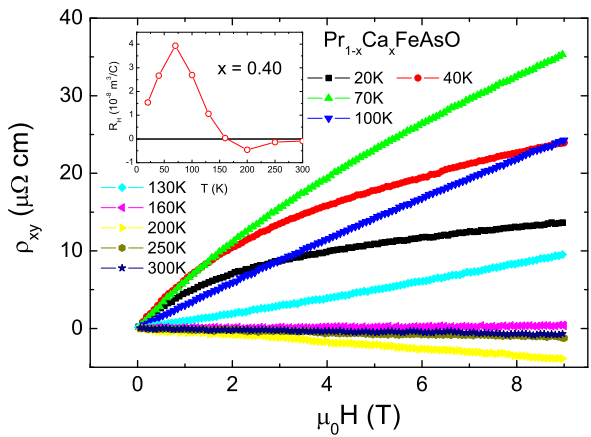}
\caption{(Color online) Hall effect measurements for one sample
Pr$_{0.60}$Ca$_{0.40}$FeAsO. The main frame shows the field
dependence of the Hall resistivity $\rho_{xy}$ at different
temperatures. Inset: Temperature dependence of the Hall coefficient
$R_H$, which is positive in the temperature region below about 160
K.} \label{fig5}
\end{figure}

\subsection{The case in the calcium-doped Pr$_{1-x}$Ca$_{x}$FeAsO}

One may be curious to know what would happen if we substitute
calcium to the systems based on other rare-earth elements. Actually,
we have tried the case of calcium-doped LaFeAsO, PrFeAsO, NdFeAsO,
GdFeAsO, and so on. Here we just show the results of
Pr$_{1-x}$Ca$_{x}$FeAsO for example. No superconductivity was found
in the calcium-doped Pr$_{1-x}$Ca$_{x}$FeAsO samples in quite wide
doping range ($0.10\leq x \leq 0.50$). In Fig. 4, we show the
temperature dependence of resistivity for two selected samples with
$x = 0.20$ and 0.40. In order to have a comparison, we also display
the resistivity data for two strontium-doped Pr$_{1-x}$Sr$_{x}$FeAsO
samples with different doping levels ($x$ = 0.05 and 0.25). It is
clear that the resistivity anomaly from the AF or structural
transition around 160 K is suppressed gradually with the increase of
strontium contents. By having a closer scrutiny, we find that the
behavior of the calcium-doped samples is between that of the two
strontium-doped samples in high temperature region. This may suggest
that the real doped charge carriers in the calcium-doped PrFeAsO are
roughly corresponding to $0.10 \sim0.20$ of that of the
strontium-doped PrFeAsO system and the AF order has been suppressed
to a certain extent by the calcium-doping.

To further investigate the conducting properties of the
Pr$_{1-x}$Ca$_{x}$FeAsO sample, we have measured the Hall effect of
them and display the data of one typical sample with $x = 0.40$ in
Fig. 5. Nonlinear behavior was observed in the field dependent
$\rho_{xy}$ data below 100 K. The hump feature and positive value of
$R_H$ can be seen below about 160 K, which is rather similar to that
observed in Tb$_{1-x}$Ca$_{x}$FeAsO (see Fig. 3). This behavior
indicates strongly that the hole-type charge carriers have been
induced to the system and they dominate the conducting in low
temperature region.

\begin{table}[p]
\caption{Selected Rietveld refinements results of the calcium-doped
Pr$_{0.87}$Ca$_{0.13}$FeAsO (0.13 is the value from the refinements)
at room temperature, along with that of the parent phase PrFeAsO and
strontium-doped Pr$_{0.84}$Sr$_{0.16}$FeAsO for comparison. The data
of the latter two samples are cited from other
reports.\cite{Jeitschko,Ju} }
\begin{tabular}
{ccccccc}\hline \hline
sample & $a$ ($\AA$)  &  $c$ ($\AA$)   & Fe-As-Fe ($^o$)  & d$_{PrOPr}$ ($\AA$)   & d$_{AsFeAs}$ ($\AA$)  & d$_{inter}$ ($\AA$)\\
\hline
PrFeAsO                      & $3.985$  \vline  & $8.595$   & $111.968$    & $2.405$  & $2.690$  & $1.750$  \\
Pr$_{0.87}$Ca$_{0.13}$FeAsO  & $3.987$  \vline  & $8.628$   & $112.429$    & $2.396$  & $2.668$  & $1.782$  \\
Pr$_{0.84}$Sr$_{0.16}$FeAsO  & $3.985$  \vline  & $8.622$   & $112.413$    & $2.387$  & $2.666$  & $1.785$  \\
 \hline \hline
\end{tabular}
\label{tab:table1}
\end{table}

In order to find the factors which prevent Pr$_{1-x}$Ca$_{x}$FeAsO
from superconducting even if hole-type charge carriers have been
doped into the system, we analyzed the structural details of one
calcium-doped Pr$_{1-x}$Ca$_x$FeAsO with nominal $x$ = 0.4. Selected
Rietveld refinement results are listed in table I, and the
refinement pattern is shown in Fig. 6. Only small amounts of
Pr$_2$O$_3$ and FeAs can be seen as the impurities. From the
refinement, we find that the actual doping concentration for the
sample Pr$_{1-x}$Ca$_x$FeAsO with nominal $x$ = 0.4 is only about
0.13, which is quite consistent with the argument we obtained from
the resistivity data (see Fig. 4). In order to have a comparison
with the strontium-doped PrFeAsO system where superconductivity has
been obtained, we also display the structural parameters of the
parent phase PrFeAsO and strontium-doped Pr$_{1-x}$Sr$_x$FeAsO with
$x$ = 0.16 (the value from refinement) \cite{Jeitschko,Ju} in table
I. Here we define d$_{PrOPr}$ and d$_{AsFeAs}$ as the vertical
distance between the Pr atoms residing at the top and bottom of the
PrO layer and that between the As atoms in the FeAs layer,
respectively. And d$_{inter}$ is the interlayer space between the
Pr-O-Pr block and the As-Fe-As block. It is clear that the lattice
constant along $a$-axis remains nearly unchanged while that along
$c$-axis expands clearly when calcium is doped to PrFeAsO. Both
d$_{PrOPr}$ and d$_{AsFeAs}$ shrink slightly while d$_{inter}$
expands distinctly with the actual calcium doping of 0.13, resulting
in the expansion behavior of the lattice along $c$-axis.
Surprisingly, we find that calcium doping gives a rather similar
influence to the crystal structure to that of strontium doping, when
we compare the parameters of Pr$_{0.87}$Ca$_{0.13}$FeAsO and
Pr$_{0.84}$Sr$_{0.16}$FeAsO as shown in table I, even if the radius
of Ca$^{2+}$ is smaller than that of Pr$^{3+}$ while the radius of
Sr$^{2+}$ is larger (see Fig. 7). We note that the sample
Pr$_{0.84}$Sr$_{0.16}$FeAsO still doesn't superconduct and
superconductivity was achieved in the samples with even higher
doping, as reported in Ref.[25]. However, the fact that the radius
of Ca$^{2+}$ is smaller than that of Pr$^{3+}$ seems to prevent from
doping even more calcium to PrFeAsO and consequently prevent from
achieving superconductivity in Pr$_{1-x}$Ca$_{x}$FeAsO system,
because the effect of doped calcium is to expand the lattice along
$c$-axis. This argument is reinforced by the fact that only 13\% of
calcium can be doped into the system even if the nominal doping
concentration is 40\%.

\begin{figure}
\includegraphics[width=9cm]{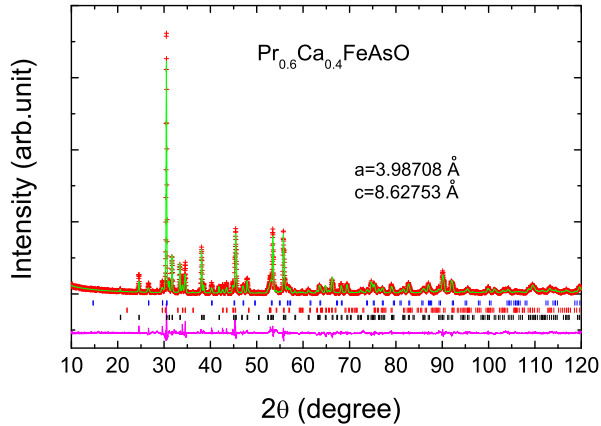}
\caption{(Color online) The observed (red crosses) and calculated
(green solid line) x-ray powder diffraction patterns of
Pr$_{0.6}$Ca$_{0.4}$FeAsO. The three rows of vertical bars show the
calculated positions of Bragg reflections for Pr$_2$O$_3$ (blue),
FeAs(red) and Pr$_{0.6}$Ca$_{0.4}$FeAsO (black), respectively. The
magenta solid line shown at the bottom of the figure indicates the
differences between observations and calculations.} \label{fig6}
\end{figure}

\begin{figure}
\includegraphics[width=9cm]{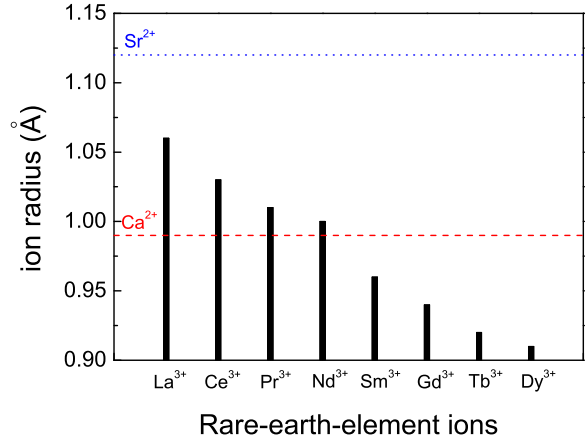}
\caption{(Color online) The data of ion radii for some selected
rare-earth-element ions. The blue dotted and red dashed lines
represent the value of Sr$^{2+}$ and Ca$^{2+}$, respectively.}
\label{fig7}
\end{figure}

To validate our supposition, we have tried the case of other
calcium-doped samples. In La$_{1-x}$Ca$_{x}$FeAsO, we found that it
is quite difficult to dope charge carriers (probably also Ca) to the
system and the Hall coefficient remains negative, and very similar
behaviors to that of Pr$_{1-x}$Ca$_{x}$FeAsO were observed in
Nd$_{1-x}$Ca$_{x}$FeAsO. Very small superconducting signal can be
observed in Sm$_{1-x}$Ca$_{x}$FeAsO sometimes. Zero resistance can
be obtained easily in Gd$_{1-x}$Ca$_{x}$FeAsO, but the diamagnetic
signal is smaller than that of the Tb$_{1-x}$Ca$_{x}$FeAsO system.
We can't get the 1111 phase on the heavy rare-earth (Dy, Ho, et al)
side under ambient pressure. By summarizing the phenomena mentioned
above, we argue that the relationship between the ion radii of the
rare-earth elements and the alkaline-earth element may play a key
role in achieving superconductivity in the hole-doped 1111 phase
(see Fig. 7). That is, superconductivity emerges only when the ion
radius of the rare-earth element is smaller than that of the
alkaline-earth element. At this time, we can say safely that it is
rather difficult, if not impossible, to obtain superconductivity
when the ion radius of the rare-earth element is larger. Moreover,
it seems that superconductivity favors the situation when the
difference between the two radii is lager, within the tolerance of
crystal lattice. These arguments are quite consistent with that
stated in the previous paragraph. This actually gives a restriction
in exploring new superconductors in hole-dope side of 1111 phase.

\section{Concluding remarks}

In summary, bulk superconductivity was achieved by substituting
Tb$^{3+}$ with Ca$^{2+}$ in TbFeAsO system. The maximum
superconducting transition temperature $T_c$ = 15.6 K is found to
appear around the nominal doping level $x$ = 0.40$\sim$ 0.50. The
positive Hall coefficient $R_H$ in a wide low-temperature range
suggests that the hole-type charge carriers dominate the conduction
in this system. Surprisingly, the slope of the upper critical
magnetic field vs. temperature near $T_c$ in calcium-doped
Tb$_{1-x}$Ca$_{x}$FeAsO is found to be much higher than that of the
electron-doped and strontium-doped ones. Moreover, we have
investigated the structural and conducting properties of other
calcium-doped systems (taking Pr$_{1-x}$Ca$_{x}$FeAsO for example).
We found that the relationship between the ion radii of the
rare-earth elements and alkaline-earth elements may play a key role
in achieveing superconductivity in the hole-doped 1111 phase.

\begin{acknowledgments}
We acknowledge the help of XRD experiments from L. H. Yang and H.
Chen. This work is supported by the Natural Science Foundation of
China, the Ministry of Science and Technology of China (973 project:
2006CB01000, 2006CB921802), the Knowledge Innovation Project of
Chinese Academy of Sciences (ITSNEM).
\end{acknowledgments}

\end{document}